\documentclass[runningheads]{llncs}
\RequirePackage[colorlinks=true,allcolors=blue!70!black]{hyperref}
\RequirePackage[numbers]{natbib}

\RequirePackage{changepage}
\RequirePackage{footmisc}
\RequirePackage[export]{adjustbox}
\RequirePackage{amsmath,amsfonts,amssymb,paralist}
\RequirePackage[normalem]{ulem}
\RequirePackage[noend]{algpseudocode}
\RequirePackage{enumitem} \setlist{nosep}
\RequirePackage{colortbl,multirow,booktabs}
\RequirePackage[skip=0pt plus1pt]{caption}
\RequirePackage[rightcaption, raggedright]{sidecap}
\sidecaptionvpos{figure}{c} \sidecaptionvpos{table}{c}
\RequirePackage[textsize=scriptsize]{todonotes}

\makeatletter
\newcommand{\@chapapp}{\relax}
\renewcommand\@biblabel[1]{#1.}
\g@addto@macro{\normalsize}{%
    \setlength{\abovedisplayskip}{0pt}
    \setlength{\abovedisplayshortskip}{0pt}
    \setlength{\belowdisplayskip}{0pt}
    \setlength{\belowdisplayshortskip}{0pt}}
\makeatother

\dbltextfloatsep \floatsep
\textfloatsep \floatsep \intextsep \floatsep

\begin{document}
\pagestyle{plain} \thispagestyle{plain}

\def\titletext{Automated Early Leaderboard Generation From Comparative Tables}
\title{\titletext}
\author{Mayank Singh$^\P$ \and Rajdeep Sarkar$^\dagger$ \and
  Atharva Vyas$^\dagger$ \and Pawan Goyal$^\dagger$ \and \\
  Animesh Mukherjee$^\dagger$ \and Soumen Chakrabarti$^\ddag$}
\institute{IIT Gandhinagar$^\P$, IIT~Kharagpur$^\dagger$, and IIT~Bombay$^\ddag$}
\authorrunning{Singh et al.}
\maketitle

\begin{abstract}
A \emph{leaderboard} is a tabular presentation of performance scores of the best competing techniques that address a specific scientific problem. Manually maintained leaderboards take time to emerge, which induces a latency in performance discovery and meaningful comparison.  This can delay dissemination of best practices to non-experts and practitioners.  Regarding papers as proxies for techniques, we present a new system to automatically discover and maintain leaderboards in the form of partial orders between papers, based on performance reported therein.  In principle, a leaderboard depends on the task, data set, other experimental settings, and the choice of performance metrics.  Often there are also tradeoffs between different metrics.  Thus, leaderboard discovery is not just a matter of accurately extracting performance numbers and comparing them.  In fact, the levels of noise and uncertainty around performance comparisons are so large that reliable traditional extraction is infeasible.  We mitigate these challenges by using relatively cleaner, structured parts of the papers, e.g., performance tables.  We propose a novel \emph{performance improvement graph} with papers as nodes, where edges encode noisy performance comparison information extracted from tables.  Every individual performance edge is extracted from a table with citations to other papers. These extractions resemble (noisy) outcomes of `matches' in an incomplete tournament.  We propose several approaches to rank papers from these noisy `match' outcomes.  We show that our ranking scheme can reproduce various manually curated leaderboards very well. Using widely-used lists of state-of-the-art papers in 27 areas of Computer Science, we demonstrate that our system produces very reliable rankings.  We also show that commercial scholarly search systems cannot be used for leaderboard discovery, because of their emphasis on citations, which favors classic papers over recent performance breakthroughs.  Our code and data sets will be placed in the public domain.
\end{abstract}

\section{Introduction}
\label{sec:Intro}

Comparison against best prior art is critical for publishing experimental research.  With the explosion of online research paper repositories like arXiv, and the frenetic level of activity in some research areas, keeping track of the best techniques and their reported performance on benchmark tasks has become increasingly challenging. \textit{Leaderboards}, a tabular representation of the performance scores of some of the most competitive techniques to solve a scientific task, are now commonplace. However, most of these leaderboards are manually curated and therefore take time to emerge.  The resulting latency presents a barrier to entry of new researchers and ideas, trapping ``wisdom'' about winning techniques to small coteries, disseminated by word of mouth. Thus, automatic leaderboard generation is an interesting research challenge. Recent work \citep{hashimoto7automatic} has focused on automatic synthesis of reviews from multiple scientific documents.  However, to the best of our knowledge, no existing system incorporates \emph{comparative experimental performance} reported in papers into the process of leaderboard generation.

\noindent \textbf{Limitations of conventional information extraction:} The ordering of competing techniques in a leaderboard depends on a large number of factors, including the task being solved, the data set(s) used, sampling protocols, experimental conditions such as hyperparameters, and the choice of performance metrics.  Further, there are often tradeoffs between various competing metrics, such as recall vs.\ precision, or space vs.\ time.  In fact, an accurate extraction, in conjunction with all the contextual details listed above, is almost impossible.  We argue that conventional table and quantity extraction \citep{CafarellaHWWZ2008WebTables2, SarawagiC2014QEWT} is  neither practical, nor sufficient, for leaderboard induction.  In fact, numeric data is often presented as combinations of comparative charts and tables embedded together in a single figure \citep{SinghSVGMC2019TabCite}.  These may even use subplots with multicolor bars representing baseline and proposed approaches.
  
\noindent \textbf{Table citations:}
A practical way to work around the difficult extraction problem is to focus on the relatively cleaner and more structured parts of a paper, viz., tables.  Performance numbers are very commonly presented in tables.  A prototypical performance table is shown in \citet[Figure~\ref{fig:CompTable}]{SinghSVGMC2019TabCite}.  Each row shows the name of a competing system or algorithm, along with a citation.  (A transposed table style is easily identified with simple rules.)  Each subsequent column is dedicated to some performance \textbf{metric}.  The rows make it simple to associate performance numbers with specific papers.  In recent years, tables with citations (here, named \textbf{table citations}) and performance summaries have become extremely popular in arXiv.


\noindent \textbf{Performance improvement graphs:}
We digest a multitude of tables in different papers into a novel \textbf{performance improvement graph}.  Each edge represents an instance of comparison between two papers, labeled with the ID of the paper where the comparison is reported, the metric (e.g., recall, precision, F1 score, etc.) used for the comparison, and the numeric values of the metric in the two papers. Note that every individual performance edge is extracted from a table with citations to other papers.  Each such extracted edge is noisy.  Apart from the challenge of extracting quantities from tables and recognizing their numeric types \cite{CafarellaHWWZ2008WebTables2,SarawagiC2014QEWT}, there is no control on the metric names, as they come from an open vocabulary (i.e., the column headers are arbitrary strings).  Processing one table is a form of `micro' reading; we must aggregate these `micro' readings into a satisfactory `macro' reading comparing two papers.  We propose several reasonable edge aggregation strategies to simplify and featurize the performance improvement graph, in preparation for ranking papers.

\noindent \textbf{Ranking papers using table citation tournaments:}
Ranking sports teams into total orders, on the basis of the win/loss outcomes of a limited number of matches played between them, has a long history \cite{Jech1983Tournament,David1987Tournament,Redmond2003Tournament}.  We adapt two widely-used tournament solvers and find that they are better than some simple baselines.  However, we can further improve on tournament solvers using simple variations of PageRank \cite{PageBMW1998pagerank,xing2004weighted} on a graph suitably derived from the tournament.  Overall, our best ranking algorithms are able to produce high-quality leaderboards that agree very well with various manually curated leaderboards.  In addition, using a popular list of papers spanning 27 different areas of Computer Science, we show that our system is able to produce reliable rankings of the state-of-the-art papers. We also demonstrate that commercial academic search systems like Google Scholar (GS)\footnote{\protect\url{https://scholar.google.com/}} and Semantic Scholar (SS)\footnote{\protect\url{https://semanticscholar.org/}} cannot be used (and, in fact, are not intended to be used) for discovering leaderboards, because of their emphasis on aggregate citations, which typically favors classic papers over latest performance leaders.

\section{Emergence of leaderboards}
\label{leaderboard}

Experts in an area are usually familiar with latest approaches and their performance.  In contrast, new members of the community and practitioners need guidance to identify the best-performing techniques.  This gap is currently bridged by ``organically emerging'' leaderboards that organize and publish the names and the performance scores of the best algorithms in a tabular form.  Such leaderboards are commonplace in Computer Science, and in many other applied sciences.

The prime limitation of manually curated leaderboards is the natural latency until the performance numbers in a freshly-published paper are noticed, verified, and assimilated.  This can induce delays in the dissemination of the best techniques to non-experts.  In this paper, we build an end-to-end system to automate the process of leaderboard generation.  The system is able to mine table citations, extract noisy performance comparisons from these table citations, aggregate the micro readings to a smooth macro reading and finally obtain rankings of papers.

\begin{table}[ht]
 \caption{Ability of GS, SS, and our system to recall prominent leaderboard papers for the PASCAL VOC Challenge.} \label{eg_leaderboard}
\resizebox{\hsize}{!}{
\begin{tabular}{lccc} \toprule
Paper&GS&SS&Our\\\midrule
\normalsize Encoder-Decoder with Atrous Separable Convolution for Semantic Image Segmentation&\cellcolor{red!20}$\times$&\cellcolor{red!20}$\times$&\cellcolor{red!20}$\times$\\
\normalsize Rethinking Atrous Convolution for Semantic Image Segmentation&\cellcolor{red!20}$\times$&\cellcolor{red!20}$\times$&\cellcolor{green!20}$\checkmark$\\
\normalsize Pyramid Scene Parsing Network&\cellcolor{red!20}$\times$&\cellcolor{red!20}$\times$&\cellcolor{green!20}$\checkmark$\\
\normalsize Wider or Deeper: Revisiting the ResNet Model for Visual Recognition&\cellcolor{red!20}$\times$&\cellcolor{red!20}$\times$&\cellcolor{red!20}$\times$\\
\normalsize RefineNet: Multi-Path Refinement Networks for High-Resolution Semantic Segmentation&\cellcolor{green!20}$\checkmark$&\cellcolor{red!20}$\times$&\cellcolor{red!20}$\times$\\
\normalsize Understanding Convolution for Semantic Segmentation&\cellcolor{red!20}$\times$&\cellcolor{red!20}$\times$&\cellcolor{green!20}$\checkmark$\\
\normalsize Not All Pixels Are Equal: Difficulty-aware Semantic Segmentation via Deep Layer Cascade&\cellcolor{red!20}$\times$&\cellcolor{red!20}$\times$&\cellcolor{green!20}$\checkmark$\\
\normalsize Identifying Most Walkable Direction for Navigation in an Outdoor Environment&\cellcolor{red!20}$\times$&\cellcolor{red!20}$\times$&\cellcolor{red!20}$\times$\\
\normalsize Fast, Exact and Multi-Scale Inference for Semantic Image Segmentation with Deep \dots &\cellcolor{red!20}$\times$&\cellcolor{red!20}$\times$&\cellcolor{red!20}$\times$\\
\normalsize DeepLab: Semantic Image Segmentation with Deep Convolutional Nets, Atrous Convolution, \dots &\cellcolor{red!20}$\times$&$\cellcolor{green!20}\checkmark$&\cellcolor{green!20}$\checkmark$\\
\normalsize Laplacian Pyramid Reconstruction and Refinement for Semantic Segmentation&\cellcolor{green!20}$\checkmark$&\cellcolor{red!20}$\times$&\cellcolor{green!20}$\checkmark$\\
\normalsize High-performance Semantic Segmentation Using Very Deep Fully Convolutional Networks&\cellcolor{red!20}$\times$&\cellcolor{red!20}$\times$&\cellcolor{green!20}$\checkmark$\\
\normalsize Higher Order Conditional Random Fields in Deep Neural Networks&\cellcolor{red!20}$\times$&\cellcolor{red!20}$\times$&\cellcolor{red!20}$\times$\\
\normalsize Efficient piecewise training of deep structured models for semantic segmentation&\cellcolor{green!20}$\checkmark$&\cellcolor{green!20}$\checkmark$&\cellcolor{green!20}$\checkmark$\\
\normalsize Semantic Image Segmentation via Deep Parsing Network&\cellcolor{red!20}$\times$&\cellcolor{green!20}$\checkmark$&\cellcolor{green!20}$\checkmark$\\
\normalsize Semantic Image Segmentation with Task-Specific Edge Detection Using CNNs \dots &\cellcolor{green!20}$\checkmark$&\cellcolor{red!20}$\times$&\cellcolor{green!20}$\checkmark$\\
\normalsize Pushing the Boundaries of Boundary Detection using Deep Learning&\cellcolor{red!20}$\times$&\cellcolor{red!20}$\times$&\cellcolor{green!20}$\checkmark$\\
\normalsize Attention to Scale: Scale-aware Semantic Image Segmentation&\cellcolor{green!20}$\checkmark$&\cellcolor{green!20}$\checkmark$&\cellcolor{green!20}$\checkmark$\\
\normalsize BoxSup: Exploiting Bounding Boxes to Supervise Convolutional Networks \dots &\cellcolor{green!20}$\checkmark$&\cellcolor{green!20}$\checkmark$&\cellcolor{red!20}$\times$\\
\normalsize Learning Deconvolution Network for Semantic Segmentation&\cellcolor{green!20}$\checkmark$&\cellcolor{green!20}$\checkmark$&\cellcolor{green!20}$\checkmark$\\
\normalsize Conditional Random Fields as Recurrent Neural Networks&\cellcolor{red!20}$\times$&\cellcolor{red!20}$\times$&\cellcolor{red!20}$\times$\\
\normalsize Weakly- and Semi-Supervised Learning of a DCNN for Semantic Image Segmentation&\cellcolor{red!20}$\times$&\cellcolor{red!20}$\times$&\cellcolor{green!20}$\checkmark$\\
\normalsize Bayesian SegNet: Model Uncertainty in Deep Convolutional Encoder-Decoder Architectures \dots &\cellcolor{red!20}$\times$&\cellcolor{red!20}$\times$&\cellcolor{red!20}$\times$\\
\normalsize Semantic Image Segmentation with Deep Convolutional Nets and Fully Connected CRFs&\cellcolor{red!20}$\times$&\cellcolor{green!20}$\checkmark$&\cellcolor{green!20}$\checkmark$\\
\normalsize Global Deconvolutional Networks for Semantic Segmentation&\cellcolor{red!20}$\times$&\cellcolor{red!20}$\times$&\cellcolor{red!20}$\times$\\
\normalsize Convolutional Feature Masking for Joint Object and Stuff Segmentation&\cellcolor{red!20}$\times$&\cellcolor{red!20}$\times$&\cellcolor{red!20}$\times$\\
\bottomrule \hline
\end{tabular} }
\end{table}

In Table~\ref{eg_leaderboard}, we show an example leaderboard generated by our system (details of the system to be discussed later in the subsequent sections) for the PASCAL VOC Challenge (which involves semantic segmentation of images).  Similar results reproducing other leaderboards are presented by~\citet{SinghSVGMC2019TabCite}.  We observe that our system is able to find many of the papers present in this human-curated leaderboard.  Traditional academic search systems like GS and SS do not fare well in finding leaderboard entries; each returned only seven papers (see Table~\ref{eg_leaderboard}) in their top 50 results retrieved for the query `semantic segmentation'.  Systems that emphasize cumulative citations rather than performance scores cannot be used for leaderboard discovery.  Citations to a paper that make incremental improvements, resulting in the best experimental performance, may never catch up with the seminal paper that introduced a general problem or technique.

\section{Limits of conventional table information extraction}
\label{limitsofie}

Performance displays are implicitly connected to a complex context developed in the paper, including the task, the data set, choice of training and test folds, hyperparameters and other experimental settings, performance metrics etc.  Millions of reviewer hours are spent each year weighing experimental evidence based on the totality of the experimental context.  ``Micro-reading'' one table at a time is not likely to replace that intellectual process.   Beyond contextual ambiguities, there are often trade-offs between different metrics like space vs.\ time, recall vs.\ precision, etc.  In summary, leaderboard induction is not merely a matter of accurately extracting performance numbers and numerically comparing them.

One way to partly mitigate the above challenges is to use relatively cleaner, structured parts of the papers, e.g., single tables or single charts. We focus on tables in our first-generation system.  However, with advanced visual chart mining and OCR~\citep{mitra2006automatic,al2015automatic,Jung:2017}, we can conceivably extend the system to charts as well. 

We concentrate on (the increasing number of) tables that also cite papers, which are surrogates for techniques.  Table~\ref{tab:dist_table_cit} shows the average number of citations in a paper $p$ that occur in tables, against the year of publication of~$p$.  Clearly, there is a huge surge in the use of table citations in the last five years, which further motivates us to exploit them for building our system. 

\begin{table}[t]
    \centering
    \begin{tabular}{|l|c|c|c|c|c|c|c|c|c|c|c|c|c|}
    \hline
    Year&2005&2006&2007&2008&2009&2010&2011&2012&2013&2014&2015&2016&2017\\\hline
    Average&0.0&0.0&0.12&0.17&0.082&0.18&0.40&0.46&0.57&1.04&3.22&3.61&4.06\\\hline
    \end{tabular}
    \caption{Average number of table citations made by an arXiv paper between 2005 and 2017.}
    \label{tab:dist_table_cit}
\end{table}

\begin{figure}[t]
\centering\includegraphics[trim=0 660 0 1,clip,width=.8\hsize]{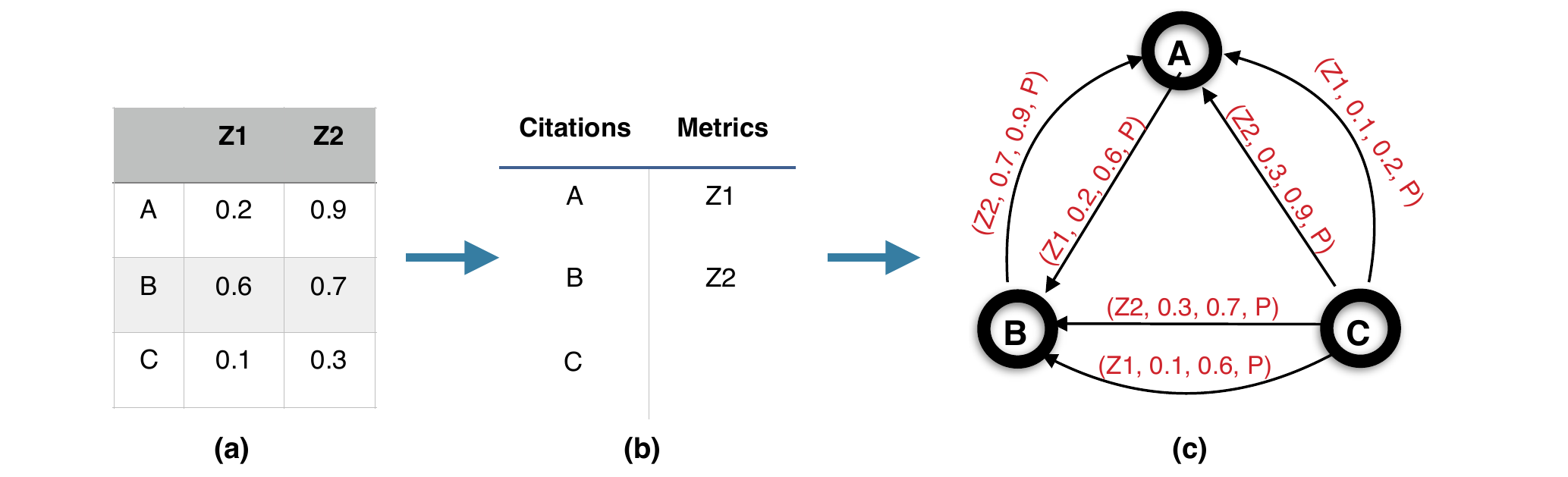}
\caption{First table extraction step toward performance tournament graph construction: (a) An example table present in paper $P$ comparing three methods, $A$, $B$ and $C$, for two evaluation metrics, $Z1$ and $Z2$. (b) Unique citations to the methods as well the evaluation metrics used are extracted, and (c) an abstract performance tournament graph is constructed. 
}
\label{fig:network_construction}
\end{figure}


\section{Performance improvement graph}
\label{pig}

\subsection{Raw performance improvement graph}

The performance improvement graph $G(V,E,Z)$ is a directed graph among a set of research papers $V$ that are compared against each other. Here,  $Z$ represents the set of all the evaluation metrics.  An edge between two papers $(A,B)$ (see Figure~\ref{fig:network_construction}) is annotated with four-tuple $(z, v_1, v_2, P)$, where $z\in Z$, $v_1$ and $v_2$ represent the metric value (`recall', `F1', `time') and lower and higher performing papers respectively.  $P$ denotes the paper that compared $A$ and $B$.  The directionality of an edge $e$ ($e\in E$) is determined by the performance comparison between two endpoints.  The paper with lower performance points toward better performing paper.  Simple heuristic rules are used to orient the edges. E.g., large F1 but small running times\footnote{`Time' is ambiguous by itself: a long time on battery but short training time are preferred. Our system is meant to take such errors it might make in stride.} are preferred. Figure~\ref{fig:network_construction} shows a toy example of the construction of a raw performance improvement graph from an extracted table. 

One table provides just one noisy comparison signal between two papers or techniques.  Although table citations allow us to make numerical comparisons, there is no guarantee of the same data set or experimental conditions across different tables, leave alone different papers.
Therefore, we process the raw performance improvement graph in two steps:
\begin{description}[leftmargin=1em]
\item[Local sanitization:]
All directed edges connecting a pair of papers in the raw performance improvement graph are replaced with one directed edge in the sanitized performance improvement graph.  This is partly a denoising step, described through the rest of this section (\ref{sec:sanitization}).
\item[Global aggregation:]
In section~\ref{minegraph}, we present and propose various methods of analyzing the sanitized performance improvement graph to arrive at a total order for the nodes (papers) to present in a synthetic leaderboard.
\end{description}

\subsection{Sanitized performance improvement graph}
\label{sec:sanitization}

\subsubsection{Relative edge improvement (REI) distribution:}
One unavoidable characteristic of the raw performance improvement graph is the existence of noisy edges from incomparable or botched extractions.  We define
\begin{equation}
\label{eq:improvement}
\text{REI}_z(u,v) = \frac{v_z - u_z}{u_z}
\end{equation}
where $(u,v)$ represents a directed edge from paper $u$ to $v$; $u_z$ and $v_z$ denote performance scores of paper $u$ and $v$ respectively against a metric~$z$. As described in previous section, $u_z$ is lower than~$v_z$.

We computed REIs from four leaderboards described in Section~\ref{dataset}. These improvement scores are computed by considering all pairs of papers present in the respective leaderboards.  We note that less than 0.5\% of the edges have REI above 100\%.  In contrast, manual inspection of various erroneously extracted edges revealed that their REI was much larger than~100\%.  Therefore, we sanitize the raw performance improvement graph by pruning edges having improvement scores larger than~100\%.  This simple thresholding yielded graphs as clean as by using supervised learning (details omitted) to remove noisy edges.

\subsubsection{Sanitizing multi-edges:}
Every comparison creates a directed edge with different tuple value.  A pair of papers can be compared in multiple tables, resulting in (anti-) parallel edges or multi-edges.  Two directed edges are termed as \emph{anti-parallel} if they are between the same pair of papers, but in opposite directions.  Whereas, two directed edges are said to be \emph{parallel} if they are between the same pair of papers and in the same direction.  In Figure~\ref{fig:network_construction}c, two parallel edges exist between papers $B$ and $C$ and two anti-parallel edges exist between papers $A$ and~$B$.

Multiple strategies can be utilized to summarize and aggregate multi-edges into a condensed tournament graph.  We consider the following variations. Note that all of these are directed graphs.  In each case, we discuss if and how a directed edge $(i,j)$ is assigned a summarized weight.

\noindent\textbf{UNW --- Unweighted Graph:}
The simplest variant preserves the directed edges without any weights. This is equivalent to giving a weight of 1 for each of these directed edge $(i,j)$, if there is any comparison.

\noindent\textbf{ALL --- Weighted graph (total number of comparisons):} 
This variation uses the total number of comparisons between two papers $p_i$ and $p_j$ as the weights of the directed edge. Thus, each time an improvement is reported, it is used as an additional vote to obtain the edge weight.

\noindent\textbf{SIG -- Sigmoid of actual improvements on edges:}
This variation takes into account the sigmoid value of the actual improvement score. If paper $u$ having a score of $u_z$ on a specific metric $z$, improves upon paper $v$ which has a score of $v_z$ in the same table and same metric, we compute the improvement score using Eq.~\eqref{eq:improvement}. We then pass this score through a sigmoid function of the form: 
\begin{equation} 
\sigma_z(u,v) = \frac{1}{1+e^{-\textrm{REI}_z(u,v)}}
\end{equation}
To combine the multiple improvement scores of $u$ over $v$ on different metrics and, thereby, obtain the edge weights, we use the following two techniques.
\begin{description}[leftmargin=.5em]
\item[Max:] We set the weight of the edge pointing from $v$ to $u$ as the maximum of all the sigmoid values of the improvement scores across the different metrics. 
\item[Average:] We set the weight of the edge pointing from $v$ to $u$ as the average of all the sigmoid values of the improvement scores across the different metrics.
\end{description}

\subsubsection{Dummy winner and loser nodes:}
In the tournament ranking literature that we shall discuss in the next section, the most prominent factor that guarantees convergence is that the tournament must be connected. However, performance tournament graphs are mostly disconnected due to extraction inaccuracies, incomplete article collection, etc.  Therefore, we introduce a dummy node that either wins or loses over all other nodes in the graph. A dummy node has a suitably directed edge to every other node.

\section{Mining sanitized performance improvement graphs}
\label{minegraph}

In this section, we explore several ranking schemes to select the most competitive papers by analyzing the sanitized performance graph. We begin with basic baselines, then explore and adapt the tournament literature, and finally present adaptations of PageRank-style algorithms. Solving an incomplete tournament over $n$ teams means to assign each team a score or rank inducing a total order over them, and presents a natural analogy with incomplete pairwise observations.  The literature on tournaments seeks to extrapolate the anticipated outcome of a match between teams $i$ and $j$ (which was never played, say) in terms of the statistics of known outcomes, e.g., $i$ defeated~$k$ and $k$ defeated~$j$.

\noindent \textbf{Sink nodes:}
We can ignore the numeric values in table cells and regard each table as comparing some papers, a pair at a time, and inserting an edge from paper $p_1$ to paper $p_2$ if the table lists a better (greater or smaller depending on metric) number against $p_2$ than $p_1$.  In such a directed graph, sink nodes that have no out-links are locally maximal.  Thus, the hunt for leaders may be characterized as a hunt for sink nodes.  We do not expect this to work well, because our graphs contain many biconnected components, thanks to papers being compared on multiple metrics.

\noindent \textbf{Cocitation:}
An indirect indication that a paper has pushed the envelope of performance on a task is that it is later compared with many papers.  We can capture this signal in a graph where nodes are papers, and an edge and its reverse edge (both unweighted) are added between papers $p_1$ and $p_2$ if they are cited by any paper.  Edges in both directions are added without considering the numbers extracted from the tables.

\noindent \textbf{Linear tournament:}
As described earlier, incomplete tournament presents a natural analogy to performance comparisons. \cite{Redmond2003Tournament} started with an incomplete tournament matrix $M$ where $m_{ij} = m_{ji}$ is the number of matches played between teams $i$ and~$j$.  
$\pmb{m}=(m_i)$ where $m_i = \sum_j m_{ij}$ is the number of matches played by team~$i$.  Abusing the division operator, let $\bar{M} = M/\pmb{m}$ denote $M$ after normalizing rows to add up to~1.

Of the $m_{ij}$ matches between teams $i$ and $j$, suppose $i$ won $r_{ij}$ times and $j$ won $r_{ji} = m_{ij} - r_{ij}$ times.  Then the \emph{dominance} of $i$ over $j$ is $d_{ij} = r_{ij} - r_{ji}$ and the dominance of $j$ over $i$ is $d_{ji} = r_{ji}-r_{ij} = -d_{ij}$.  Setting the dominance of a team over itself as zero in one dummy match, we can calculate the average dominance of a team $i$ as $\bar{d}_i = \left[\sum_j d_{ij}\right]\left/\left[\sum_{j} m_{ij}\right]\right.$, and this produces a reasonable ranking of the teams to a first approximation, i.e., up to ``first generation'' or direct matches.  To extrapolate to ``second generation'' matches, we consider all $(i,k)$ and $(k,j)$ matches, which is given by the matrix $M^2$.  Third generation matches are likewise counted in $M^3$, and so on.  \cite{David1987Tournament} showed that a meaningful scoring of teams can be obtained as the limit $\lim_{T\to\infty}\sum_{t=0}^T \bar{M}^t \cdot \bar{\pmb{d}}$, where $\bar{\pmb{d}}=(\bar{d}_i)$.

\noindent \textbf{Exponential tournament:}
The exponential tournament model \cite{Jech1983Tournament} is somewhat different, and based on a probabilistic model.  Given $R = (r_{ij})$ as above, it computes row sums $\rho_i = \sum_j r_{ij}$.  Let $\pmb{\rho} = (\rho_i)$ be the empirically observed team scores.  Again, we can sort teams by decreasing $\rho_i$ as an initial estimate, but this is based on an incomplete and noisy tournament.  Between teams $i$ and $j$ there are (latent/unknown) probabilities $p_{ij} + p_{ji} = 1$ such that the probability that $i$ defeats $j$ in a match is $p_{ij}$.  Then the MLE estimate is $p_{ij} = r_{ij}/m_{ij}$.  \cite{Jech1983Tournament} shows that there exist team `values' $\pmb{v} = (v_i)$ such that $\sum_i v_i = 0$ and
\begin{align}
\rho_i &= \sum_j m_{ij} p_{ij} = \sum_j \frac{m_{ij}}{1 + \exp(v_j - v_i)}.
\end{align}
Here $M$ and $\pmb{\rho}$ are observed and fixed, and $\pmb{v}$ are variables.
Values $\pmb{v}$ can be fitted using gradient descent.  Once the matrix $\pmb{P}=(p_{ij})$ is thus built, it gives a consistent probability for all possible permutations of the teams.  In particular, $\prod_j p_{ij}$ gives the probability that $i$ defeats all other teams (marginalized over all orders within the other teams $j$).  Sorting teams $i$ by decreasing $\prod_j p_{ij}$ is thus a reasonable rating scheme.

\noindent \textbf{PageRank:}
PageRank computes a ranking of the competitive papers in the (suitably aggregated) tournament graph based on the structure of the incoming links. We utilize standard PageRank implementation\footnote{https://networkx.github.io} to rank nodes in the directed weighted tournament graph. We found best results (see Table~\ref{tab:comparison}) when damping factor ($\alpha$) is set at 0.90. We run this weighted variant of PageRank on each induced tournament graph corresponding to each query. The induced tournament graph consists of papers ($P$) relevant to the query along with the papers  compared with $P$. These candidate response papers are ordered using $PR$ values. These scores can also be used for tie-breaking sink nodes.

\section{Experiments}
\label{experiments}

\subsection{Datasets}
\label{dataset}

\subsubsection{\textit{ArXiv} dataset:}
We downloaded (in June 2017) the entire arXiv document source dump but
restricted this study to the field of Computer Science. 
Table~\ref{tab:arxiv_cs_dataset} shows statistics of arXiv's Computer Science papers.  ArXiv 
mandates uploading the source of DVI, PS, or PDF articles generated from \LaTeX\ code resulting in a large volume of papers (1,181,349 out of 1,297,992 papers) with source code.

\begin{table}[t]
\centering
  \caption{Salient statistics about the arXiv and Computer Science data sets.}
  \label{tab:arxiv_cs_dataset}
  \resizebox{0.8\textwidth}{!}{
  \begin{tabular}{llcccllc} \cline{1-3}\cline{6-8}
   \parbox[t]{2mm}{\multirow{4}{*}{\rotatebox[origin=c]{90}{\textbf{Full}}}}&Year range& 1991--2017&&&\parbox[t]{2mm}{\multirow{7}{*}{\rotatebox[origin=c]{90}{\textbf{Comp. Science}}}}&Number of papers &107,795\\ 
  &papers &1,297,992&&&&Year range& 1993--2017  \\
  &papers with \LaTeX\ code & 1,181,349 &&&&Total references& 2,841,554\\
  &Total fields& 9&&&&Total indexed papers& 1,145,083\\\cline{1-3}
  &&&&&&Total tables& 204,264\\
  &&&&&&Total table citations & 98,943\\
  &&&&&&Unique extracted metrics & 14,947\\
  \cline{6-8}
 \end{tabular}}
\end{table}
 
\noindent \textbf{Preprocessing and extracting table citations:}
The curation process involves several sub-tasks such as reference extraction, reference mapping, table extraction, collecting table citations, performance metrics extraction and edge orientation.  Due to space constraints, we present detailed description and evaluation of each sub-tasks elsewhere~\citep{SinghSVGMC2019TabCite}.

\subsubsection{State-of-the-art deep learning  papers:}
A representative example from the rapidly growing and evolving area of deep learning is {\small \url{https://github.com/sbrugman/deep-learning-papers}}. The website contains state-of-the-art (SOTA) papers on malware detection/security, code generation, NLP tasks like summarization, classification, sentiment analysis etc., as well as computer vision tasks like style transfer, image segmentation, and self-driving cars. This Github repository is very popular and has more than 2,600 stargazers and has been forked 330 times. The repository notes 27 different popular topics shown in Table~\ref{tab:motivation_1}. The table also shows that the SOTA papers curated by knowledgeable experts rarely find a place in the top results returned by the two popular academic search systems --- GS and SS.  To be fair, these systems were not tuned to find SOTA papers, but we argue that this is an important missing search feature.  As fields saturate and stabilize, citations to ``the last of the SOTA papers'' may eclipse citations to older ones, rendering citation-biased ranking satisfactory.  But we again argue that recognizing SOTA papers quickly is critical to researchers, especially new comers and practitioners.

\subsubsection{Organic leaderboards:}
We identify manually curated leaderboards that compare competitive papers on specific tasks. The four popular leaderboards that we choose for our subsequent experiments are (i) The Stanford Question Answering Dataset (SQuAD)\footnote{https://rajpurkar.github.io/SQuAD-explorer/}, (ii) Pixel-Level Semantic Labeling Task (Cityscapes)\footnote{https://www.cityscapes-dataset.com/benchmarks/}, (iii) VOC Challenge (PASCAL)\footnote{https://goo.gl/6xTWxB}, and (iv) MIT Saliency (MIT-300)\footnote{http://saliency.mit.edu/results\_mit300.html}. 
Each leaderboard consists of several competitive papers compared against multiple metrics. For example, the SQuAD leaderboard consists of 117 competitive papers compared against two metrics `Exact Match' and `F1 score'.  
The tasks mostly include topics from natural language processing (e.g., question answering) and image processing (e.g., semantic labeling, image segmentation and saliency prediction).

\begin{table}[t]
\centering
\caption{Recall of human-curated state-of-the-art (SOTA) deep learning papers within top-10 and top-20 responses from two popular academic search engines (Google Scholar and Semantic Scholar).  Both systems show low visibility of SOTA papers.} \label{tab:motivation_1}
\resizebox{\hsize}{!}{
\begin{tabular}{cclllllllllllllllllllllllllllc} \toprule
&&\rotatebox[origin=l]{90}{Code Generation}&
\rotatebox[origin=l]{90}{Malware Detection}&
\rotatebox[origin=l]{90}{Summarization}&
\rotatebox[origin=l]{90}{Taskbots}&
\rotatebox[origin=l]{90}{Text Classification}&
\rotatebox[origin=l]{90}{Question Answering}&
\rotatebox[origin=l]{90}{Sentiment Analysis}&
\rotatebox[origin=l]{90}{Machine Translation}&
\rotatebox[origin=l]{90}{Chatbots}&
\rotatebox[origin=l]{90}{Reasoning}&
\rotatebox[origin=l]{90}{Gaming}&
\rotatebox[origin=l]{90}{Style Transfer}&
\rotatebox[origin=l]{90}{Object Tracking}&
\rotatebox[origin=l]{90}{Visual Q\&A}&
\rotatebox[origin=l]{90}{Image Segmentation}&
\rotatebox[origin=l]{90}{Text Recognition}&
\rotatebox[origin=l]{90}{Brain Comp. Interfacing}&
\rotatebox[origin=l]{90}{Self Driving Cars}&
\rotatebox[origin=l]{90}{Object Recognition}&
\rotatebox[origin=l]{90}{Logo Recognition}&
\rotatebox[origin=l]{90}{Super Resolution}&
\rotatebox[origin=l]{90}{Pose Estimation}&
\rotatebox[origin=l]{90}{Image Captioning}&
\rotatebox[origin=l]{90}{Image Compression}&
\rotatebox[origin=l]{90}{Image Synthesis}&
\rotatebox[origin=l]{90}{Face Recognition}&
\rotatebox[origin=l]{90}{Audio Synthesis}&\rotatebox[origin=l]{90}{\textbf{Total}}\\\hline
\#SOTA&&7&3&3&2&15&1&2&6&2&1&14&6&1&1&15&6&3&2&30&4&5&4&9&1&9&8&6&166\\\hline
\multirow{2}{*}{GS}&Top-10&0&0&0&0&0&0&0&1&0&0&0&1&0&1&0&0&0&1&1&0&0&0&1&0&0&0&0&6(3.6\%)\\\cline{2-2}
&Top-20&0&0&0&0&1&0&0&1&0&0&0&3&0&1&1&1&0&1&1&0&0&0&1&0&0&0&1&12 (7.2\%)\\\hline
\multirow{2}{*}{SS}&Top-10&0&0&0&0&0&0&0&0&0&0&0&2&0&1&0&0&0&1&1&0&0&0&1&0&0&1&0&7 (4.2\%)\\\cline{2-2}
&Top-20&0&0&0&0&0&0&0&1&0&0&0&2&0&1&1&1&0&1&1&0&1&0&1&0&0&1&0&11 (6.6\%)\\
\bottomrule \hline
\end{tabular}}
\end{table}


\subsection{Ranking state-of-the-art papers}
\label{sec:rank_SOTA}

Table~\ref{tab:comparison} shows comparisons between Google Scholar (GS), Semantic Scholar (SS), and several ranking variations implemented in our testbed.  Recall@10, Recall@20, NDCG@10, and NDCG@20 are used as the evaluation measures, averaged over the 27 topics shown in Table~\ref{tab:motivation_1}. Since our primary objective is to find competitive prior art, recall is more important in case of Web search, where precision at the top (NDCG) is paid more attention.

Given the complex nature of performance tournament ranking, our absolute recall and NDCG are modest. Among naive baselines, sink node search led to generally worst performance, which was expected. The numeric comparison is slightly better, but not much.

GS and SS are mediocre as well. Despite the obvious fit between our problem and tournament algorithms, they are surprisingly lackluster.  In fact, many of the tournament variants lose to simple cocitation. PageRank on unweighted improvement graphs performs beyond cocitation. However, the ``sigmoid'' versions of PageRank improve upon the unweighted case, almost doubling the gains beyond GS and SS, and are clearly the best choice.

\begin{table}[ht]
\centering
\caption{Comparison between several ranking schemes. Recall@10, Recall@20, NDCG@10, NDCG@20 measures are averaged over the 27 tasks (queries). OS: Online Systems; LT: Linear Tournament; ET: Exponential Tournament; ALL: Weighted graph (total number of comparisons); UNW: Unweighted directed performance graph; SIG: Sigmoid of the actual performance improvement; DW: Dummy Winner; DL: Dummy Loser, DCC: Dense co-citation, NC: Numeric comparison.} \label{tab:comparison}
\resizebox{\hsize}{!}{
\begin{tabular}{cccccccccccccccccccccccc} \toprule
&&\multicolumn{2}{c}{OS}&&\multicolumn{4}{c}{LT}&&\multicolumn{4}{c}{ET}&&\multicolumn{4}{c}{PageRank}&&\multicolumn{1}{c}{Sink}&&\multicolumn{2}{c}{BS}\\\cline{3-4}\cline{6-9}\cline{11-14}\cline{16-19}\cline{21-21}\cline{23-24}
&&\rotatebox[origin=l]{90}{GS}&
\rotatebox[origin=l]{90}{SS}&&
\rotatebox[origin=l]{90}{DW}&
\rotatebox[origin=l]{90}{DL}&
\rotatebox[origin=l]{90}{DW}&
\rotatebox[origin=l]{90}{DL}&&
\rotatebox[origin=l]{90}{DW}&
\rotatebox[origin=l]{90}{DL}&
\rotatebox[origin=l]{90}{DW}&
\rotatebox[origin=l]{90}{DL}&&
\rotatebox[origin=l]{90}{UNW}&
\rotatebox[origin=l]{90}{ALL}&
\rotatebox[origin=l]{90}{Avg.}&
\rotatebox[origin=l]{90}{Max.}&&
\rotatebox[origin=l]{90}{ALL}&&
\rotatebox[origin=l]{90}{DCC}&
\rotatebox[origin=l]{90}{NC}\\
&&&&&\multicolumn{2}{c}{ALL}&\multicolumn{2}{c}{SIG}&&\multicolumn{2}{c}{ALL}&\multicolumn{2}{c}{SIG}&&&&\multicolumn{2}{c}{SIG}&&&&&\\\hline
\multirow{2}{*}{\rotatebox[origin=c]{90}{T-10}}&Recall \%&\cellcolor{pink}7.38&\cellcolor{pink}7.84&&4.63&4.63&1.8&1.93&&1.7&2.31&1.7&1.7&&\cellcolor{green!20}19.35&\cellcolor{green!20}16.86&\cellcolor{green!20}19.35&\cellcolor{green!20}19.35&&0.62&&\cellcolor{yellow!20}12.91&\cellcolor{yellow!20}6.73\\
&NDCG&\cellcolor{pink}0.073&\cellcolor{pink}0.065&&0.029&0.029&0.016&0.019&&0.027&0.024&0.02&0.02&&\cellcolor{green!20}0.151&\cellcolor{green!20}0.131&\cellcolor{green!20}0.154&\cellcolor{green!20}0.149&&0.009&&\cellcolor{yellow!20}0.142&\cellcolor{yellow!20}0.036\\\hline
\multirow{2}{*}{\rotatebox[origin=c]{90}{T-20}}&Recall \%&\cellcolor{pink}10.48&\cellcolor{pink}10.08&&5.86&5.86&6.5&6.63&&4.17&2.93&2.93&2.93&&\cellcolor{green!20}21.74&\cellcolor{green!20}21.95&\cellcolor{green!20}22.36&\cellcolor{green!20}22.09&&0.62&&\cellcolor{yellow!20}19.25&\cellcolor{yellow!20}7.35\\
&NDCG&\cellcolor{pink}0.086&\cellcolor{pink}0.074&&0.034&0.034&0.028&0.03&&0.036&0.026&0.025&0.025&&\cellcolor{green!20}0.159&\cellcolor{green!20}0.151&\cellcolor{green!20}0.164&\cellcolor{green!20}0.159&&0.009&&\cellcolor{yellow!20}0.152&\cellcolor{yellow!20}0.037\\
\bottomrule
\end{tabular}}
\end{table}

\begin{table}[th]
  \caption{Recall@50 and NDCG@50 measures for four leaderboards. Green cells indicate best scores and red cells indicate worst scores.}
  \label{tab:leaderboard_recall_ndcg}
\resizebox{\hsize}{!}{
  \begin{tabular}{ccccccccccccccc} \toprule
\multirow{2}{*}{Leaderboard name} & \multicolumn{2}{c}{GS}&&\multicolumn{2}{c}{SS}
&& \multicolumn{2}{c}{PageRank UNW}&& \multicolumn{2}{c}{PageRank SIG (Avg)}&& \multicolumn{2}{c}{PageRank SIG (Max)}\\ \cline{2-3}\cline{5-6}\cline{8-9}\cline{11-12} \cline{14-15}
&Recall (\%) &NDCG&&Recall (\%)&NDCG&&Recall (\%)&NDCG&&Recall (\%)&NDCG&&Recall (\%)&NDCG\\\midrule
SQuAD&\cellcolor{red!20}0&\cellcolor{red!20}0&&7.14&0.014&&\cellcolor{green!20}21.42&\cellcolor{green!20}0.206&&\cellcolor{green!20}21.42&0.205&&14.29&0.177\\
Cityscapes&\cellcolor{red!20}25&\cellcolor{red!20}0.067&&37.5&0.159&&\cellcolor{green!20}62.5&0.303&&\cellcolor{green!20}62.5&\cellcolor{green!20}0.310&&\cellcolor{green!20}62.5&0.295\\
PASCAL&\cellcolor{red!20}26.92&\cellcolor{red!20}0.12&&\cellcolor{red!20}26.92&0.179&&\cellcolor{green!20}57.69&0.497&&\cellcolor{green!20}57.69&0.500&&\cellcolor{green!20}57.69&\cellcolor{green!20}0.502\\
MIT-300&42.86&0.115&&\cellcolor{red!20}14.28&\cellcolor{red!20}0.036&&\cellcolor{green!20}50.00&\cellcolor{green!20}0.465&&\cellcolor{green!20}50.00&0.437&&\cellcolor{green!20}50.00&0.438\\
\bottomrule \hline
 \end{tabular}}
\end{table}

\subsection{Leaderboard generation}

In this section, we demonstrate our system's capability to automatically generate task-specific leaderboards. We utilize four manually curated leaderboards for this study. Automatic leaderboard generation procedure is divided into two phases:\\
\noindent \textbf{Obtaining list of candidate papers relevant to a task:}
We, first, obtain a list of candidate papers relevant to a given task. We utilize textual information such as title and abstract to find relevant candidate papers. These candidate papers are further ranked by utilizing best performing PageRank schemes (described in section~\ref{sec:rank_SOTA}). We consider top-50 ranked results and show comparisons between Google Scholar (GS), Semantic Scholar (SS), and top-3 high performing PageRank variations against two evaluation measures --- Recall@50 and NDCG@50 ---  in Table~\ref{tab:leaderboard_recall_ndcg}. As expected, GS and SS performed poorly for all of the four leaderboards. PageRank variations have almost double the gains beyond GS and SS and are clearly the best choice.   Some generated leaderboards are listed in~\citet{SinghSVGMC2019TabCite}.

\noindent \textbf{Ranking candidate papers to generate leaderboard:}
Next, we compute the correlation between ranks in generated leaderboards with the ground-truth ranks obtained from the organic leaderboards. Table~\ref{tab:leaderboard_scores} presents the Spearman's rank correlation of rankings produced by PageRank variations, UNW, SIG (Avg) and SIG (Max), with the corresponding ground-truth rankings for the four leaderboards.  SQuAD shows the highest correlation (0.94 for F1 and 0.89 for EM) for all of the three PageRank variations.  CityScapes and PASCAL also exhibit impressive correlation coefficients for all the PageRank variants. For the MIT-300 leaderboard, while the correlation coefficient is decent for the SIM metric it is somewhat low for the AUC metric. The reason for the low correlation is existence of multiple weakly connected components.  A local winner in one component is affecting the global ranks across all components. 

\begin{SCtable}
\resizebox{0.5\textwidth}{!}{
\begin{tabular}{lclcccc} \toprule
Name&Nodes&Metric&UNW&SIG (AVG)&SIG (MAX) \\\midrule
\multirow{2}{*}{SQuAD}&\multirow{2}{*}{9}&F1&0.94&0.94&0.94\\
&&EM&0.89&0.89&0.89\\ \hline
CityScapes&7&iIoU&0.7&0.7&0.7\\ \hline
PASCAL&26&AP&0.57&0.57&0.57\\ \hline
\multirow{2}{*}{MIT-300}&\multirow{2}{*}{9}&AUC&0.23&0.23&0.23\\
&&SIM&0.53&0.45&0.45\\
\bottomrule \hline
 \end{tabular}}
\caption{Spearman's rank correlation of rankings produced by UNW, SIG (Avg) and SIG (Max) with the corresponding ground-truth rankings for the four leaderboards.}
\label{tab:leaderboard_scores}
\end{SCtable}

\subsection{Effect of graph sanitization}

As described in section~\ref{sec:sanitization}, graph sanitization is a necessary preprocessing step. In this section, we present several real examples that resulted in greater visibility of state-of-the-art after sanitization. As representative examples, we consider two tasks, ``image segmentation'' and ``gaming'', to show how graph sanitization results in noise reduction in the performance improvement graphs. We find several state-of-the-art papers that performed poorer than a competitive paper with high improvement score ($>$700\%). This anomaly resulted in the poorer visibility of the state-of-the-art papers in top ranks. However, after sanitization, the visibility gets improved. For example, Table~\ref{tab:sanitization_effect} shows four examples of high improvement edges whose removal resulted in the higher recall of the state-of-the-art papers. 

\begin{table}[h]
\centering
  \caption{Effect of graph sanitization. The first two edges correspond to the task of ``image segmentation'' and the last two to the task of ``gaming''. Removal of these edges resulted in higher visibility of SOTA papers.}
  \label{tab:sanitization_effect}
  \resizebox{0.5\textwidth}{!}{
  \begin{tabular}{cccc} \toprule
  Source&Destination&Improvement \%&Back-edge (Y/N) \\\midrule  
1511.07122&1504.01013&775&Y\\
1511.07122&1511.00561&6597&Y\\\hline
1611.02205&1207.4708&4012.3&N\\
1412.6564&1511.06410&928.8&N\\
\bottomrule \hline
 \end{tabular}}
\end{table}

\subsection{Why is PageRank better than tournaments?}

PageRank variations performed significantly better than tournament variations. Several assumptions of tournament literature do not hold true for scientific performance graphs; for instance, existence of disconnected components is a common characteristic of performance graphs. Unequal number of comparisons between a pair of papers in performance graphs is another characteristic that demarcates it from the tournament settings. We observe that in a majority of task-specific performance graphs, tournament-based ranking scheme is biased toward papers with zero out-degrees. Therefore the tournaments mostly converge to the global sinks; in fact, we observe more than half of the tournament based top-ranked papers are sink nodes.  This is why recall and NDCG in Table~\ref{tab:comparison} for these two methods are close.

\section{Conclusion and future scope}
\label{sec:end}
We introduce performance improvement graphs that encode information about performance comparisons between scientific papers.  The process of extracting tournaments is designed to be robust, flexible, and domain-independent, but this makes our labeled tournament graphs rather noisy.  We present a number of ways to aggregate the tournament edges and a number of ways to score and rank nodes on the basis of this incomplete and noisy information.  
In ongoing work, we are extending beyond \LaTeX\ tables to line, bar and pie charts \citep{mitra2006automatic, al2015automatic}.

\noindent\paragraph*{\bfseries Acknowledgment:}
Partly supported by grants from IBM and Amazon.

\bibliographystyle{abbrvnat} 
\bibliography{voila,tournament}

\clearpage
\appendix



\title{\titletext \\ (Supplementary Material)}
\author{Mayank Singh$^\P$ \and Rajdeep Sarkar$^\dagger$ \and
  Atharva Vyas$^\dagger$ \and Pawan Goyal$^\dagger$ \and \\
  Animesh Mukherjee$^\dagger$ \and Soumen Chakrabarti$^\ddag$}
\institute{IIT Gandhinagar$^\P$, IIT~Kharagpur$^\dagger$, and IIT~Bombay$^\ddag$}
\authorrunning{Singh et al.}
\titlerunning{Early leaderboard detection}
\pagestyle{plain} \thispagestyle{plain}
\maketitle

\begin{figure}
\begin{adjustwidth}{-2cm}{-2cm}
\centering
  \includegraphics[width=\linewidth,frame]{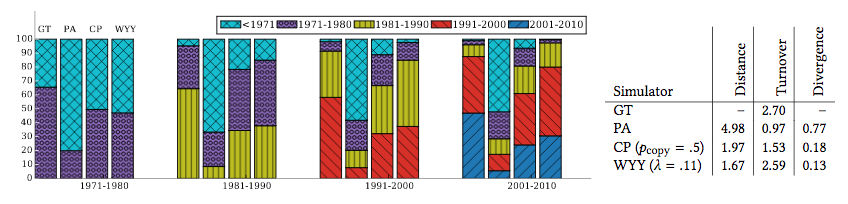}
  \caption{Comparative charts and tables embedded together in a single figure.}
  \label{fig:CompChartsTabs}
\end{adjustwidth}
\end{figure}

\begin{figure}
\begin{adjustwidth}{-2cm}{-2cm}
\centering
  \includegraphics[trim={10mm 3mm 6mm 9mm}, clip, width=\linewidth,frame]{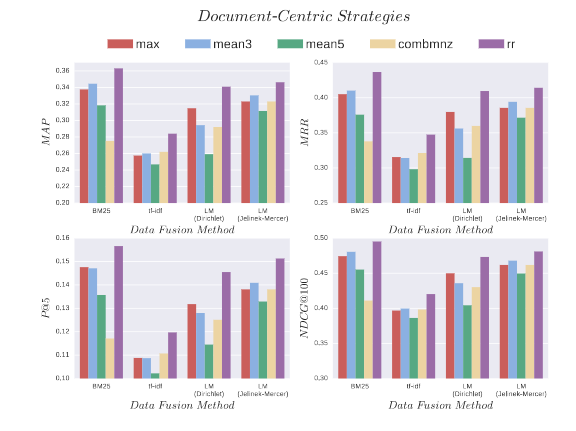}
  \caption{Multiple comparative subplots with multi-color bars representing baseline papers.}
  \label{fig:CompSubplots}
\end{adjustwidth}
\end{figure}

\begin{figure}
\begin{adjustwidth}{-2cm}{-2cm}
   \centering
  \includegraphics[width=\linewidth]{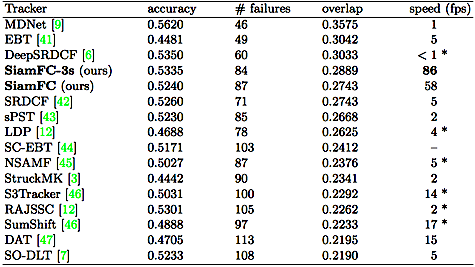}
  \caption{Sample performance numbers in a table with citations.  Each
    row corresponds to a competing algorithm or system, which is
    associated with a paper cited (green highlighted link) from that
    row.  Each column represents a performance metric.}
  \label{fig:CompTable}
\end{adjustwidth}
\end{figure}

\begin{table}
\begin{adjustwidth}{-2cm}{-2cm}
\centering
\resizebox{\linewidth}{!}{
\begin{tabular}{|l|c|c|c|c|c|}
\hline
\multicolumn{1}{|c|}{\multirow{2}{*}{Paper}} & \multirow{2}{*}{GS} & \multirow{2}{*}{SS} & \multicolumn{2}{c|}{SIG}& \multirow{2}{*}{UNW}\\\cline{4-5}
&&&MAX&AVG&\\\hline
Reinforced Mnemonic Reader for Machine Reading Comprehension &
\cellcolor{red!20}$\times$ & \cellcolor{red!20}$\times$ & \cellcolor{red!20}$\times$ & \cellcolor{red!20}$\times$ & \cellcolor{red!20}$\times$ \\
Structural Embedding of Syntactic Trees for Machine Comprehension &
\cellcolor{red!20}$\times$ & \cellcolor{red!20}$\times$ & \cellcolor{red!20}$\times$ & \cellcolor{red!20}$\times$ & \cellcolor{red!20}$\times$ \\
ReasoNet: Learning to Stop Reading in Machine Comprehension &
\cellcolor{red!20}$\times$ & \cellcolor{red!20}$\times$ & \cellcolor{red!20}$\times$ & \cellcolor{green!20}$\checkmark$ & \cellcolor{green!20}$\checkmark$ \\
Bidirectional Attention Flow for Machine Comprehension                                     & \cellcolor{red!20}$\times$                   & \cellcolor{red!20}$\times$                   & \cellcolor{red!20}$\times$     & \cellcolor{red!20}$\times$     & \cellcolor{red!20}$\times$     \\
Multi-Perspective Context Matching for Machine Comprehension                               & \cellcolor{red!20}$\times$                   & \cellcolor{red!20}$\times$                   & \cellcolor{red!20}$\times$     & \cellcolor{red!20}$\times$     & \cellcolor{red!20}$\times$     \\
Exploring Question Understanding and Adaptation in Neural-Network-Based \dots & \cellcolor{red!20}$\times$                   & \cellcolor{red!20}$\times$                   & \cellcolor{green!20}$\checkmark$ & \cellcolor{green!20}$\checkmark$ & \cellcolor{green!20}$\checkmark$ \\
Dynamic Coattention Networks For Question Answering                                        & \cellcolor{red!20}$\times$                   & $\cellcolor{green!20}$\checkmark               & \cellcolor{red!20}$\times$     & \cellcolor{red!20}$\times$     & \cellcolor{red!20}$\times$     \\
Ruminating Reader: Reasoning with Gated Multi-Hop Attention                                & \cellcolor{red!20}$\times$                   & \cellcolor{red!20}$\times$                   & \cellcolor{red!20}$\times$     & \cellcolor{red!20}$\times$     & \cellcolor{red!20}$\times$     \\
Reading Wikipedia to Answer Open-Domain Question                                           & \cellcolor{red!20}$\times$                   & \cellcolor{red!20}$\times$                   & \cellcolor{red!20}$\times$     & \cellcolor{red!20}$\times$     & \cellcolor{red!20}$\times$     \\
Making Neural QA as Simple as Possible but not Simpler                                     & \cellcolor{red!20}$\times$                   & \cellcolor{red!20}$\times$                   & \cellcolor{red!20}$\times$     & \cellcolor{red!20}$\times$     & \cellcolor{red!20}$\times$     \\
Learning Recurrent Span Representations for Extractive Question Answering                  & \cellcolor{red!20}$\times$                   & \cellcolor{red!20}$\times$                   & \cellcolor{green!20}$\checkmark$ & \cellcolor{green!20}$\checkmark$ & \cellcolor{green!20}$\checkmark$ \\
Machine Comprehension Using Match-LSTM and Answer Pointer                                  & \cellcolor{red!20}$\times$                   & \cellcolor{red!20}$\times$                   & \cellcolor{red!20}$\times$     & \cellcolor{red!20}$\times$     & \cellcolor{red!20}$\times$     \\
Words or Characters? Fine-grained Gating for Reading Comprehension & \cellcolor{red!20}$\times$ & \cellcolor{red!20}$\times$                   & \cellcolor{red!20}$\times$     & \cellcolor{red!20}$\times$     & \cellcolor{red!20}$\times$ \\
End-to-End Answer Chunk Extraction and Ranking for Reading Comprehension   & \cellcolor{red!20}$\times$  & \cellcolor{red!20}$\times$                   & \cellcolor{red!20}$\times$     & \cellcolor{red!20}$\times$     & \cellcolor{red!20}$\times$ \\
\hline
\end{tabular}}
\caption{Ability of GS, SS, and our system to recall prominent leaderboard papers for the SQuAD.}
\end{adjustwidth}
\end{table}

\begin{table}
\begin{adjustwidth}{-2cm}{-2cm}
\resizebox{\linewidth}{!}{
\begin{tabular}{|l|c|c|c|c|c|}
\hline
\multicolumn{1}{|c|}{\multirow{2}{*}{Paper}} & \multirow{2}{*}{GS} & \multirow{2}{*}{SS} & \multicolumn{2}{c|}{SIG}& \multirow{2}{*}{UNW}\\\cline{4-5}
&&&MAX&AVG&\\\hline
Rethinking Atrous Convolution for Semantic Image Segmentation                                                   & \cellcolor{red!20}$\times$     & \cellcolor{red!20}$\times$     & \cellcolor{green!20}$\checkmark$ & \cellcolor{green!20}$\checkmark$ & \cellcolor{green!20}$\checkmark$ \\
Wider or Deeper: Revisiting the ResNet Model for Visual Recognition                                             & \cellcolor{red!20}$\times$     & \cellcolor{red!20}$\times$     & \cellcolor{red!20}$\times$     & \cellcolor{red!20}$\times$     & \cellcolor{red!20}$\times$     \\
RefineNet: Multi-Path Refinement Networks for High-Resolution Semantic Segmentation                             & \cellcolor{green!20}$\checkmark$ & \cellcolor{red!20}$\times$     & \cellcolor{red!20}$\times$     & \cellcolor{red!20}$\times$     & \cellcolor{red!20}$\times$     \\
Full-Resolution Residual Networks for Semantic Segmentation in Street Scenes                                    & \cellcolor{red!20}$\times$     & \cellcolor{red!20}$\times$     & \cellcolor{green!20}$\checkmark$ & \cellcolor{green!20}$\checkmark$ & \cellcolor{green!20}$\checkmark$ \\
Multi-level Contextual RNNs with Attention Model for Scene Labeling                                             & \cellcolor{red!20}$\times$     & \cellcolor{red!20}$\times$     & \cellcolor{red!20}$\times$     & \cellcolor{red!20}$\times$     & \cellcolor{red!20}$\times$     \\
DeepLab: Semantic Image Segmentation with Deep Convolutional Nets, Atrous Convolution \dots & \cellcolor{red!20}$\times$     & \cellcolor{green!20}$\checkmark$ & \cellcolor{green!20}$\checkmark$ & \cellcolor{green!20}$\checkmark$ & \cellcolor{green!20}$\checkmark$ \\
Efficient piecewise training of deep structured models for semantic segmentation                                & \cellcolor{green!20}$\checkmark$ & \cellcolor{green!20}$\checkmark$ & \cellcolor{green!20}$\checkmark$ & \cellcolor{green!20}$\checkmark$ & \cellcolor{green!20}$\checkmark$ \\
SegNet: A Deep Convolutional Encoder-Decoder Architecture for Image Segmentation                                & \cellcolor{red!20}$\times$     & \cellcolor{green!20}$\checkmark$ & \cellcolor{green!20}$\checkmark$ & \cellcolor{green!20}$\checkmark$ & \cellcolor{green!20}$\checkmark$\\\hline 
\end{tabular}}
\caption{Ability of GS, SS, and our system to recall prominent leaderboard papers for the CityScape.}
\end{adjustwidth}
\end{table}

\begin{table}
\begin{adjustwidth}{-2cm}{-2cm}
\resizebox{\linewidth}{!}{
\begin{tabular}{|l|c|c|c|c|c|}
\hline
\multicolumn{1}{|c|}{\multirow{2}{*}{Paper}} & \multirow{2}{*}{GS} & \multirow{2}{*}{SS} & \multicolumn{2}{c|}{SIG}& \multirow{2}{*}{UNW}\\\cline{4-5}
&&&MAX&AVG&\\\hline
DeepFix: A Fully Convolutional Neural Network for predicting Human Eye Fixations                      & \cellcolor{red!20}$\times$     & \cellcolor{green!20}$\checkmark$ & \cellcolor{red!20}$\times$     & \cellcolor{red!20}$\times$     & \cellcolor{red!20}$\times$     \\
A Deep Spatial Contextual Long-term Recurrent Convolutional Network for Saliency Detection            & \cellcolor{green!20}$\checkmark$ & \cellcolor{green!20}$\checkmark$ & \cellcolor{red!20}$\times$     & \cellcolor{red!20}$\times$     & \cellcolor{red!20}$\times$     \\
Predicting Human Eye Fixations via an LSTM-based Saliency Attentive Model                             & \cellcolor{red!20}$\times$     & \cellcolor{green!20}$\checkmark$ & \cellcolor{green!20}$\checkmark$ & \cellcolor{green!20}$\checkmark$ & \cellcolor{green!20}$\checkmark$ \\
SalGAN: Visual Saliency Prediction with Generative Adversarial Networks                               & \cellcolor{red!20}$\times$     & \cellcolor{green!20}$\checkmark$ & \cellcolor{red!20}$\times$     & \cellcolor{red!20}$\times$     & \cellcolor{red!20}$\times$     \\
A Deep Multi-Level Network for Saliency Prediction                                                    & \cellcolor{red!20}$\times$     & \cellcolor{red!20}$\times$     & \cellcolor{red!20}$\times$     & \cellcolor{red!20}$\times$     & \cellcolor{red!20}$\times$     \\
Deep Visual Attention Prediction                                                                      & \cellcolor{green!20}$\checkmark$ & \cellcolor{green!20}$\checkmark$ & \cellcolor{green!20}$\checkmark$ & \cellcolor{green!20}$\checkmark$ & \cellcolor{green!20}$\checkmark$ \\
Shallow and Deep Convolutional Networks for Saliency Prediction                                       & \cellcolor{red!20}$\times$     & \cellcolor{red!20}$\times$     & \cellcolor{red!20}$\times$     & \cellcolor{red!20}$\times$     & \cellcolor{red!20}$\times$     \\
DeepFeat: A Bottom Up and Top Down Saliency Model Based on Deep Features of \dots & \cellcolor{green!20}$\checkmark$ & \cellcolor{green!20}$\checkmark$ & \cellcolor{green!20}$\checkmark$ & \cellcolor{green!20}$\checkmark$ & \cellcolor{green!20}$\checkmark$ \\
Visual saliency detection: a Kalman filter based approach                                             & \cellcolor{green!20}$\checkmark$ & \cellcolor{green!20}$\checkmark$ & \cellcolor{green!20}$\checkmark$ & \cellcolor{green!20}$\checkmark$ & \cellcolor{green!20}$\checkmark$ \\
End-to-end Convolutional Network for Saliency Prediction                                              & \cellcolor{red!20}$\times$     & \cellcolor{green!20}$\checkmark$ & \cellcolor{green!20}$\checkmark$ & \cellcolor{green!20}$\checkmark$ & \cellcolor{green!20}$\checkmark$ \\
WEPSAM: Weakly Pre-Learnt Saliency Mode                                                               & \cellcolor{green!20}$\checkmark$ & \cellcolor{green!20}$\checkmark$ & \cellcolor{green!20}$\checkmark$ & \cellcolor{green!20}$\checkmark$ & \cellcolor{green!20}$\checkmark$ \\
Visual Language Modeling on CNN Image Representations                                                 & \cellcolor{green!20}$\checkmark$ & \cellcolor{green!20}$\checkmark$ & \cellcolor{red!20}$\times$     & \cellcolor{red!20}$\times$     & \cellcolor{red!20}$\times$     \\
Visual saliency estimation by integrating features using multiple kernel learning                     & \cellcolor{green!20}$\checkmark$ & \cellcolor{green!20}$\checkmark$ & \cellcolor{red!20}$\times$     & \cellcolor{red!20}$\times$     & \cellcolor{red!20}$\times$     \\
Saliency Detection by Forward and Backward Cues in Deep-CNNs                                          & \cellcolor{green!20}$\checkmark$ & \cellcolor{green!20}$\checkmark$ & \cellcolor{green!20}$\checkmark$ & \cellcolor{green!20}$\checkmark$ & \cellcolor{green!20}$\checkmark$\\\hline 
\end{tabular}}
\caption{Ability of GS, SS, and our system to recall prominent leaderboard papers for the MIT Saliency (MIT-300).}
\end{adjustwidth}
\end{table}

\end{document}